\documentclass[reprint,aps,floatfix,10pt,twocolumn]{revtex4-1}
\usepackage{graphics, color}
\usepackage{rotating} 
\usepackage{setspace} 
\usepackage{graphicx}% http://ctan.org/pkg/graphicx
\usepackage[english]{babel}
\usepackage{amsmath,amsfonts,amssymb,verbatim,float,mathtools}
\usepackage{physics}
\usepackage[titletoc,title]{appendix} 
\usepackage{hyperref}
\hypersetup{%
	pdfborder = {0 0 0}}
\usepackage{bbold}

\begin{document}
\title{Steady states of a driven dissipative dipolar XXZ chain}
\author{C. D. Parmee}
\affiliation{TCM Group, Cavendish Laboratory, University of Cambridge, JJ Thomson Avenue, Cambridge, CB3 0HE, UK}
\author{N. R. Cooper}
\affiliation{TCM Group, Cavendish Laboratory, University of Cambridge, JJ Thomson Avenue, Cambridge, CB3 0HE, UK}
\begin{abstract}
We study theoretically a driven dissipative one-dimensional XXZ spin$-1/2$ chain with dipole coupling and a tunable strength of the Ising and XY interaction. Within a mean-field approximation, we find a rich phase diagram with uniform, spin density wave, antiferromagnetic and oscillatory phases, as well as regions of phase bistability. We study the phase diagram of small quantum systems using exact diagonalisation, and compare the results to the mean-field theory. We find that while expectation values only capture the uniform phases of the mean-field theory, fluctuations about these expectation values give signatures of spatially non-uniform phases and bistabilities. We find these signatures for all ratios of the Ising to XY interaction, showing that they appear to be general features of spin$-1/2$ systems.
\end{abstract}
\date{\today} 
 \maketitle

\section{Introduction}
The study of quantum systems driven far from equilibrium has attracted much interest over the last few years. Although not as well understood as their equilibrium counterparts, non-equilibrium phenomena are in fact rather prevalent, as any experiment has some form of interaction with an environment, which will induce dissipative processes. Whereas these processes could be viewed as a nuisance, recent studies have shown that the interplay between an external drive and dissipation can produce exotic non-equilibrium phases such as spin density waves (SDW), antiferromagnetic phases (AFM), persistent long-time oscillations (OSC) and phase bistabilities within spin$-1/2$ systems \cite{Chan2015b,Lee2013a,Lee2011a,Wilson2016a,Qian2015,Parmee2018} and higher spin systems \cite{Qian2012a}. Therefore, understanding the long-time steady state phases that can occur in open quantum systems is an intense area of current research.

In order to capture the steady state phases in a macroscopically large system, it is common to employ a mean-field approximation, where spatial correlations between sites are ignored. While the use of the mean-field approximation is well understood in equilibrium phenomena, an open question that still remains is the validity of the mean-field approximation for dissipative systems. For thermal gases in three-dimensions, one expects the mean-field approximation to become valid as quantum fluctuations become negligible \cite{Carr2013,Sibalic2016}. However, for cold systems and/or systems in reduced dimensions, where quantum fluctuations are more important, this is not necessarily the case. There have already been many studies into the true phases and transitions for the driven dissipative spin systems. These have involved exact diagonalisation and quantum Monte Carlo wavefunction approaches for small system sizes \cite{Olmos2014,Rota2018,Lee2012a,Ates2012a}, or incorporating the use of Keldysh methods \cite{Maghrebi2016}, cluster mean-field \cite{Jin2016,Jin2018,Owen2018,Biella2018}, Gutzwiller Monte Carlo approaches \cite{Huybrechts2019}, corner-space renormalisation \cite{Rota2017} or variational approaches \cite{Weimer2015} and Matrix Product and tensor network methods \cite{Cui2015,Mascarenhas2015,Clark2016,Joshi2013a,Kshetrimayum2017,Honing2013} to study larger systems. These studies have shown that first order transitions in the mean-field approximation can become second order when quantum fluctuations are included, and that bistabilities can be lost. They have also shown that the emergence of certain phases, such as antiferromagnetic phases or long-time oscillations may not occur in low dimensional spin-$1/2$ systems. 
	
Despite the disagreements between mean-field and exact numerics of quantum systems, the mean-field approximation can still serve as an indicator of features that emerge in the full quantum system. For example, it has been shown that regions of mean-field bistability for the Ising model correspond to long spatial correlations in the full quantum model \cite{Hu2013}. Also, while bistability has not been observed in finite-sized quantum spin systems, the bistable nature of the mean-field solutions is evident in quantum trajectories of the system \cite{Wilson2016a,Lee2012a,Olmos2014,Ates2012a} and also results in a decrease in the spectral gap of the Liouvillian \cite{Wilson2016a}. Therefore, it is interesting to ask how general these features are when comparing quantum results to those within the mean-field approximation, and if there are other consequences of the mean-field results for the full quantum dynamics.

In this paper, we study a driven-dissipative XXZ model with a tunable XY and Ising interaction and dipole coupling as a function of detuning and external drive strength. The tunable Ising and XY interaction connects the Ising model in Ref. \cite{Lee2011a} and the XY model in Ref. \cite{Wilson2016a}. By calculating the nonequilibrium phase diagram at mean-field level, we find how the phases evolve between these two limits, finding the emergence of spin density waves, antiferromagnetism, temporal oscillations and bistabilities. We then analyse small quantum systems and carry out an in-depth comparison to our mean-field phase diagram. While we do not find the same phase diagram for the full quantum system, we do find strong signatures of the mean-field in the spin fluctuations which relate to bistabilities and spatial phases that arise in the mean-field results. The tunable Ising to XY interaction allows us to track these signatures between the XY and Ising limit, and show that the quantum results are general features.

The rest of this paper is organised as follows. In Sec. \ref{Model}, we describe the model, then in Sec. \ref{Mean-Field Phase Diagram}, we derive the mean-field phase diagram. In Sec. \ref{Quantum Phase Diagram} we look at the full quantum model and discuss the results in Sec. \ref{Discussion} before drawing conclusions in Sec. \ref{Conclusions}.

\section{Model}
\label{Model}
\begin{figure}	
		\hspace*{0cm}
		\includegraphics[scale=0.5,clip,angle=0]{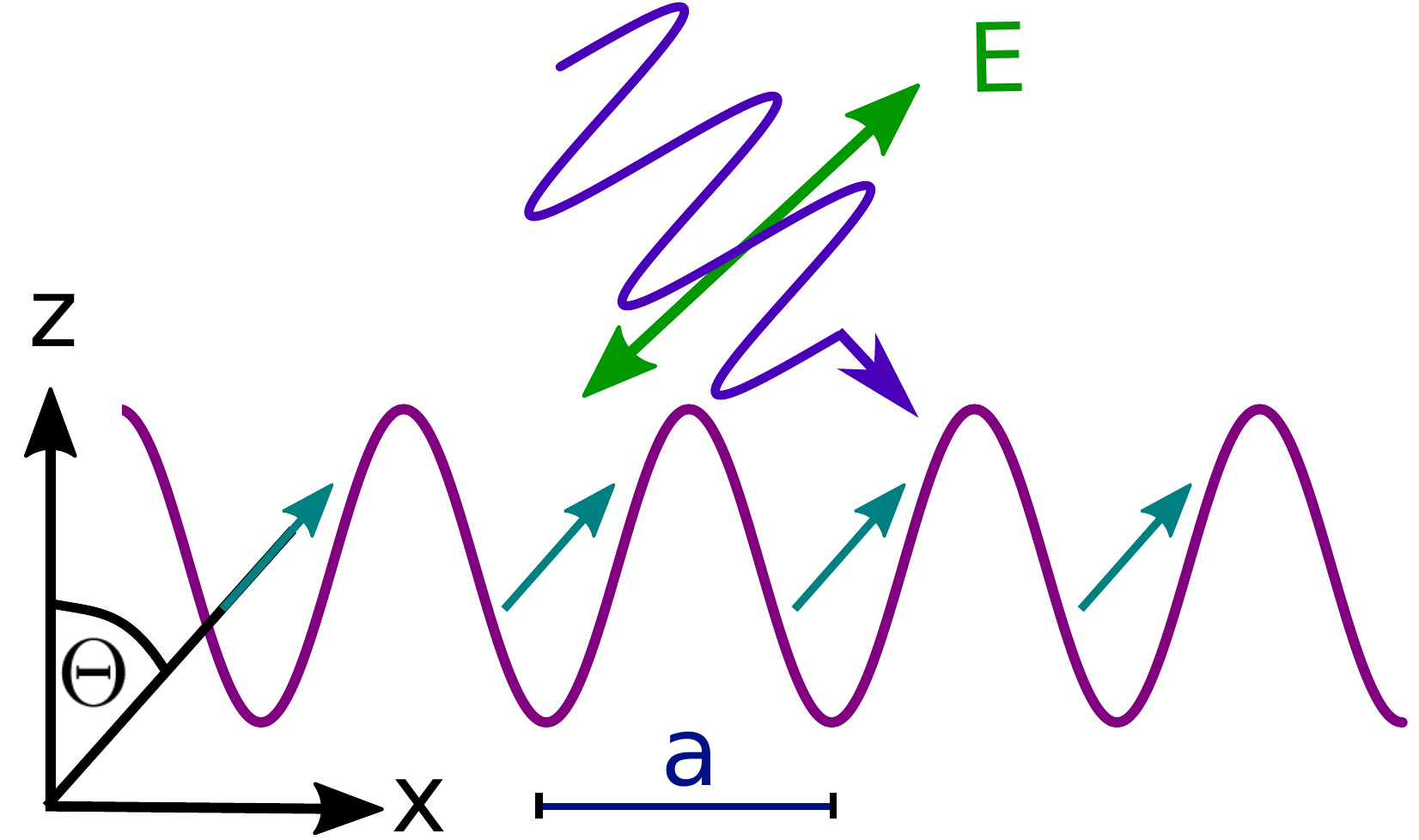}
		\caption{A schematic of a 1D array of atoms or polar molecules under an external drive. The electric field, shown by the green arrow, is oriented at an angle $\Theta$ to the z axis and controls the orientation of the dipoles. The lattice spacing is denoted by a.}
		\vspace{-0cm}
		\label{SystemModel}
	\end{figure}

The system consists of a large number, $N$, of atoms or polar molecules in a one-dimensional (1D) array. Two internal dipolar energy levels, $\ket{g}$ and $\ket{e}$, are isolated and the transition between them driven by an external drive detuned from resonance. The system can then be modelled as a spin$-1/2$ system, interacting via dipole-dipole interactions with a Rabi drive and detuning. The two-level transition also has a finite lifetime, comparable to the time scales of interaction, which results in decay from the excited state. Under the Born and Markov approximations, the dynamics of the system are described by the following master equation ($\hbar=1$)
\begin{equation}\label{MasterEq}
	\begin{split}	\frac{d\hat{\rho}(t)}{dt}=&-\text{i}\left[\hat{H},\hat{\rho}(t)\right]\\
	&+\frac{\Gamma}{2}\sum_{i}^{N}\left[2\hat{\sigma}_i^-\hat{\rho}(t) \hat{\sigma}_i^+-\left\lbrace \hat{\sigma}_i^+\hat{\sigma}^-_i,\hat{\rho}(t)\right\rbrace\right],	\end{split}
\end{equation}
where the spin operators are defined as $\hat{\sigma}^z_i=\ket{e_i}\bra{e_i}-\ket{g_i}\bra{g_i}$, $\hat{\sigma}^-_i=\ket{g_i}\bra{e_i}$ and $\hat{\sigma}^+_i=\ket{e_i}\bra{g_i}$, with $\ket{e_i}$ and $\ket{g_i}$ being the excited and ground states of the two-level transition respectively on site $i$. The decay constant, $\Gamma$, is controlled by optical pumping \cite{Lee2013a}, where the two level system is off-resonantly dressed by a third energy level. The effective decay constant of the two-level transition is then a combination of the decay rate of the original two-level system and the decay rate of the third energy level. This allows us to tune the decay independently from the interaction between the two-level systems.
The Hamiltonian is given by 
\begin{equation}\label{Hamiltonian}
\begin{split}
&\hat{H}=\frac{\Omega}{2}\sum_{i}^{N}\hat{\sigma}_{i}^{x}-\frac{\Delta}{2}\sum_{i}^{N}\hat{\sigma}_{i}^{z}\\
&+\sum_{i\ne j}^{N}\frac{J}{4R_{ij}^3}\left[\cos\alpha \hat{\sigma}_{i}^{z}\hat{\sigma}_{j}^{z}+\sin\alpha\left(\hat{\sigma}_{i}^{y}\hat{\sigma}_{j}^{y}+\hat{\sigma}_{i}^{x}\hat{\sigma}_{j}^{x}\right)\right],
\end{split}
\end{equation}
where $R_{ij}=|r_i-r_j|$, with $r_i$ being the position vector of a two-level system at site $i$, and $\hat{\sigma}^x_i=\ket{e_i}\bra{g_i}+\ket{g_i}\bra{e_i}$ and $\hat{\sigma}^y_i=\text{i}\ket{g_i}\bra{e_i}-\text{i}\ket{e_i}\bra{g_i}$. The detuning is given by $\Delta=\omega_d-\omega_{eg}$, with $\omega_d$ being the drive frequency and $\omega_{eg}$ the two-level transition frequency, and the drive strength is given by the Rabi coupling $\Omega$. We note that the power law coupling is motivated by experiment but is not crucial for the results presented here, i.e. a nearest-neighbour coupling would also work. The dipole-dipole interaction is given by $J=J_0a^3(1-3\sin^2\Theta)$ where $a$ is the lattice spacing (see Fig. \ref{SystemModel}) and the parameter $J_0$ is given by $J_0=|\textbf{d}|^2/4\pi\epsilon_0 a^3$ .
The angle $\Theta$ is the orientation of the dipoles, and can be tuned by application of a d.c. electric field. The parameter $\alpha$ relates the relative strength of the Ising and XY dipole interactions and can take values between $-\pi$ to $\pi$. However, we only focus on the range $0\leq\alpha<\pi$ as values below zero simply corresponds to a change in the sign of $J$. Tuning the value of $\alpha$ depends on the choice of internal states and external fields, with $\cos\alpha$ being related to the difference in dipole moments of the groundstate and excited state, and $\sin\alpha$ being related to the transition dipole moment between the groundstate and excited state \cite{Gorshkov2011,Peter}. 

For the remainder of the paper, we work with $\Theta=\pi/2$. Other values of $\Theta$ will result in a sign change and scaling of the interaction in 1D, but will not lead to significant changes in the types of phases that appear in our system, only in the size of the regions as a function of $\Delta$ and $\Omega$. Also, we will only study $0\leq\alpha<\pi$, as the values $-\pi\leq\alpha<0$ will result in the same phases as for $0\leq\alpha<\pi$ but with the phase diagram reflected about $\Delta=0$ due to the sign change in the interaction term. Therefore, it is sufficient to consider the range $0\leq\alpha<\pi$ and $\Theta=\pi/2$ to cover all phases that can occur in the 1D system. Finally, we choose a value of $J_0a^3/\Gamma=5$ for the rest of this paper. We should find similar results for other nearby values of $J_0a^3/\Gamma$, but expect that for $J_0a^3/\Gamma \ll 1$, we will only find spatially uniform phases in the system as the spins become effectively decoupled.

In order to find the long-time steady state of Eq. \eqref{MasterEq}, we make a Gutzwiller mean-field approximation which results in taking $\hat{\rho}(t)=\otimes \hat{\rho}_i(t)$, where $\hat{\rho}_i(t)$ is a density matrix on site $i$, and we effectively ignore correlations between spins. Then, by taking the trace of Eq. \eqref{MasterEq} over all the sites except a given site $j$, we obtain the equations of motion as
\begin{equation}\label{SpinEqs}
\begin{split}
\frac{dS_{j}^{x}}{dt}=&-\Gamma S^{x}_{j}-\Delta S^{y}_{j}\\
&+2\sin\alpha\sum_{i(\neq j)}^{N}\frac{ J}{R_{ij}^3}S_{j}^{z}S_{i}^{y}-2\cos\alpha\sum_{i(\neq j)}^{N}\frac{ J}{R_{ij}^3}S_{i}^{z}S_{j}^{y},\\
\frac{dS_{j}^{y}}{dt}=&-\Gamma S^{y}_{j}-\Omega S_{j}^{z}+\Delta S^{x}_{j}\\
&-2\sin\alpha\sum_{i(\neq j)}^{N}\frac{ J}{R_{ij}^3}S_{j}^{z}S_{i}^{x}+2\cos\alpha\sum_{i(\neq j)}^{N}\frac{ J}{R_{ij}^3}S_{i}^{z}S_{j}^{x},\\
\frac{dS_j^{z}}{dt}=&-\Gamma \left(S_{j}^{z}+\frac{1}{2}\right)+\Omega S_{j}^{y}\\
&-2\sin\alpha\sum_{i(\neq j)}^{N}\frac{ J}{R_{ij}^3}(S_{i}^{y}S_{j}^{x}-S_{j}^{y}S_{i}^{x}),
\end{split}
\end{equation}
where $S^{\beta}_j=\frac{1}{2}\Tr\left[ \hat{\sigma}^{\beta}_j\hat{\rho}(t)\right]$ are the spin expectation values. To find the steady state solutions, we solve the dynamics of the non-linear Eqs. \eqref{SpinEqs} by evolving them into the long-time limit.

\section{Mean-Field Phase Diagram}
\label{Mean-Field Phase Diagram}
\begin{figure*}
		\hspace*{0cm}
		\center
		\includegraphics[scale=0.48,clip,angle=0]{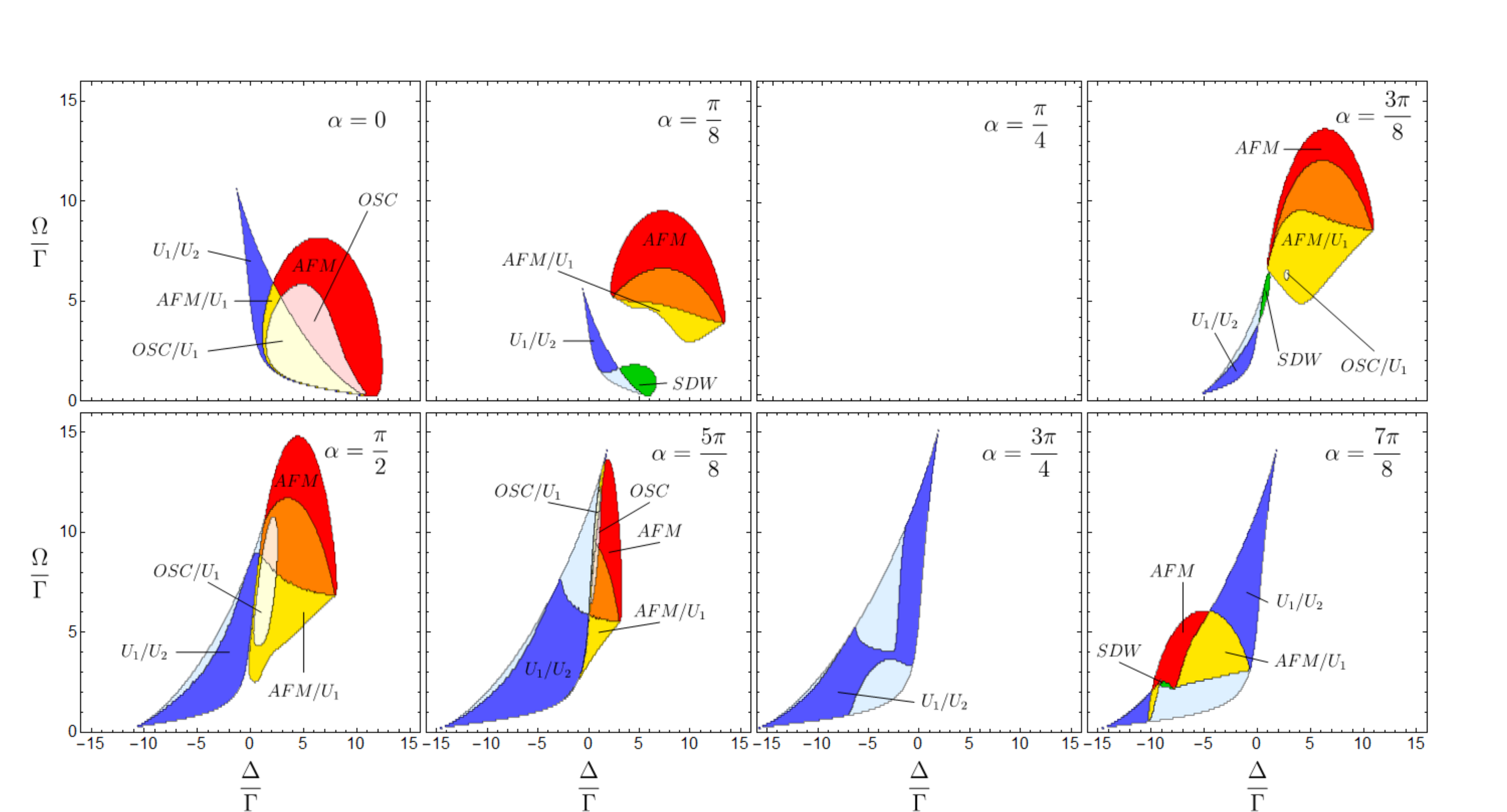}
		\caption{Phase diagrams as a function of Rabi coupling and detuning, for values of $0\leq\alpha<\pi$. We find the emergence of four key phases: uniform phases, spin density wave phases, antiferromagnetic phases and oscillatory phases. We also find examples of where these phases can be bistable with one another which means both phases can coexist within the corresponding parameter regime, and which phase the system ends up in depends on the initial conditions. These regions are denoted with double labelling e.g. AFM/$U_1$.}
		\vspace{-0cm}
		\label{phasediagrams}
	\end{figure*}

We compute the mean-field phase diagram by finding the steady states of Eqs. \eqref{SpinEqs}. To do this, we first employ a bipartite sublattice ansatz to find the uniform and antiferromagnetic steady state solutions. Depending on the parameters, there can be up to three uniform solutions and three sets of antiferromagnetic solutions. To determine the final phases that exist in the system, we perform linear stability analysis of the resultant solutions to fluctuations with wave vectors $ka = \pi m/N$, where $N$ is the number of sites on the lattice and $m$ is an integer in the range $0 \leq m \leq N$. In cases where the wave vector of instability is not equal to $0$ or $\pi$, we expect spin density wave solutions to form and for the bipartite sublattice ansatz to fail. To confirm the sublattice ansatz results are correct, we solve the full dynamics of Eqs. \eqref{SpinEqs} by evolving the equations in time until the long-time limit (up to $t\Gamma=200$) for a system size of $N=100$ with periodic boundary conditions. Simulating the full dynamics also allows us to find the resultant phases in regimes where the sublattice ansatz breaks down. For our dynamical simulations, as an initial condition, we use either $(S^x,S^y,S^z)=(0,0,-1/2)$ or, if examining phase instability, the steady state that becomes unstable. In Fig. \ref{phasediagrams}, we show a collection of phase diagrams as a function of $\Delta$ and $\Omega$ for select values of $\alpha$ in the range $0\leq\alpha<\pi$.   

The phase diagram for $\alpha=0$ is similar to a nearest-neighbour Ising model studied in Ref. \cite{Lee2011a} as expected. Similarly, the phase diagram for $\alpha=\pi/2$ is similar to a nearest-neighbour XY model studied in Ref. \cite{Wilson2016a}. However, our phase diagrams now show how these phases change as the $\alpha$ value is moved away from these two cases. For all, or almost all, $\alpha$ values, we can see some general features that occur. Specifically, we can classify four key phases that emerge in the system. Firstly, for all $\alpha$, there are the spatially uniform phases, which are shown by the white regions in the phase diagrams. At low Rabi drive, the uniform phase has a high spin magnitude and the spins close to the state with $S^z\approx-1/2$. We define this uniform phase as the $U_1$ phase. At high drive, the spin magnitude decreases and the spins lie close to the state with $S^z\approx 0$. We denote this as the $U_2$ phase. Both uniform phases occur for any spin half system even without interactions as they are just general solutions to the optical Bloch equations. For $\alpha=\pi/4$, we have a Heisenberg Hamiltonian which conserves total spin and results in only uniform phases. For all $\alpha$, the $U_1$ phase smoothly crosses over into the $U_2$ phase for most parameter ranges. However, for $\alpha \neq \pi/4$, when the drive and detuning are comparable to the interaction strength ($J/\Gamma = -5$), regimes of bistability between the $U_1$ and $U_2$ exist, which lead to sharp transitions between the two phases. Which phase the system ends up in within this region depends on the initial conditions. These regions of bistability are denoted by the dark blue regions in the phase diagram. 
	
It is also the case that when the drive and detuning are comparable to the interactions, for all $\alpha\neq \pi/4$, the uniform phases can become unstable to fluctuations, breaking translational invariance and giving rise to non-trivial phases. In the red regions, the uniform phase becomes unstable to fluctuations with a wavevector of $ka=\pi$ and a stable set of antiferromagnetic solutions exist. This results in the emergence of a canted AFM solution, with the nature of the AFM phases depending on the $\alpha$ value. The AFM solutions have the largest deviation between the $S^z$ components when $0<\alpha<\pi/4$, and a large deviation in the $S^x$ and $S^z$ components when $3\pi/4<\alpha<\pi$. However, when the XY interaction dominates, the AFM solution has the strongest deviation in the $S^y$ components. As well as instabilities to $ka=\pi$, the uniform phases can become unstable to $ka<\pi$ which results in the emergence of a SDW phase. In the SDW phase, shown by the green regions, the spin orientation varies periodically through the lattice with a period set by the instability wavevector, $ka$. We find that there are no SDW instabilities for $\alpha=0$. Therefore, it seems that SDW instabilities are related to presence of the XY interaction. The final key phase to emerge are persistent long-time oscillations, denoted by OSC, where the effects of the drive dominate over the effects from dissipation. We find that all the oscillations emerge from the instability of the AFM phase, which undergoes a Hopf bifurcation, and inherit an AFM nature. The oscillations occur in the pink regions of the phase diagrams. 
	
In several regions of the phase diagrams, multiple solutions to Eqs. \eqref{SpinEqs} coexist, which can lead to bistabilities. In the yellow regions, a stable uniform solution and stable set of AFM solutions exist, which results in AFM/$U_1$ bistability. Similarly, in the light yellow regions, there is an OSC phase which is also bistable to the $U_1$ phase. We do find that there are cases where the oscillations become unstable and so only the uniform phase exists, which we have not marked in our phase diagram. In the light blue regions, both the $U_1$ phase and $U_2$ phase exist, but one becomes unstable to fluctuations. When simulating the dynamics in these regions, we find there is predominantly only one stable uniform solution and only small regions of $U_2$/SDW phase bistability. Similarly, in the orange regions, there exists both a uniform phase and an antiferromagnetic phase, but the uniform phase is unstable. We find when simulating the dynamics in this region that only the AFM phase exists and that the AFM phase is frustrated, with the amount of frustration depending on the initial conditions. Likewise, in the light orange regions, there is an OSC phase and an unstable uniform phase. We find only the OSC phase exists, and again that there are frustration effects. 
 
\section{Quantum Phase Diagram}
\label{Quantum Phase Diagram}
Having now calculated the phase diagram at mean-field level, we now examine how it compares to the phase diagram of the full quantum system. To do this, we look at the steady state of small quantum systems with $N=8$ spins and periodic boundary conditions. Despite the small system size, we can see some distinct features emerge for the quantum system that reflect the mean-field phases. 

We first examine the spin expectation values on each site for a direct comparison to the mean-field results. We find that for the quantum system, the expectation values on each site are uniform, with no spatial variation, for all values of detuning and drive. Even for small systems, we would expect some non-uniform phases to emerge within the mean-field approximation, as the wave vectors that cause instability of the uniform phases, such as $ka=\pi$, are permitted. However, we find only uniform phases for the small quantum system, which we would expect to persist until large system sizes, due to the translational invariance of the system. If we compare the quantum expectation values to the mean-field expectation values from our phase diagrams in the regions where stable uniform phases exist, we find there is good agreement when the magnitude of the Rabi drive and detuning are large, for all $\alpha$. The difference between the expectation values becomes larger in regions where there is $U_1$/$U_2$ bistability. This is because we find no bistability in small quantum systems, but a smooth crossover between the $U_1$ and $U_2$ phases. Therefore the quantum expectation values will eventually differ from either choice of the $U_1$ or $U_2$ mean-field solution. For $\alpha=\pi/4$, we find exact agreement between the quantum and mean-field expectation values for all detuning and Rabi drive due to the Heisenberg symmetry. 

\begin{figure}
\hspace*{0cm}
\includegraphics[scale=0.45,clip,angle=0]{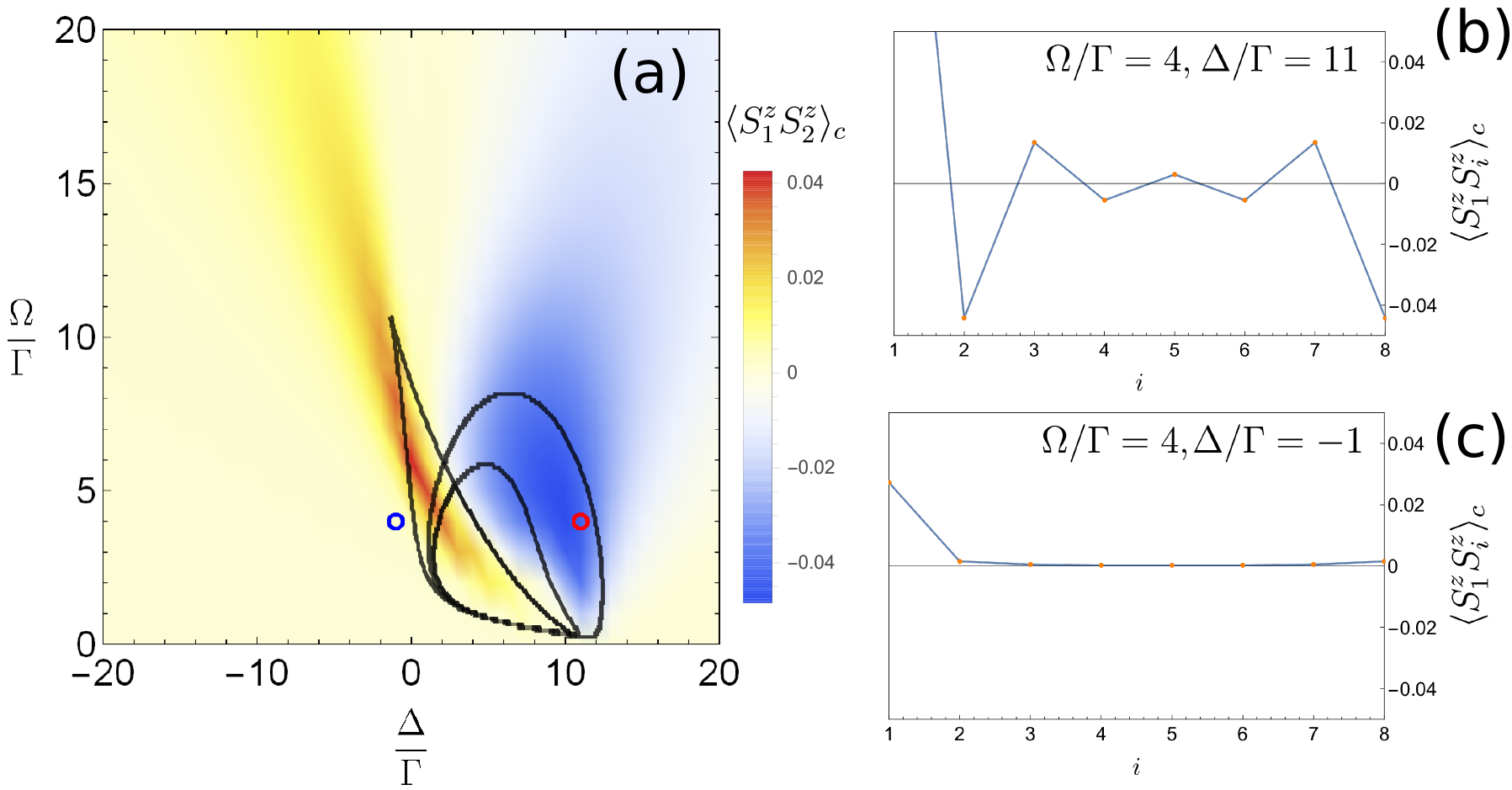}
\caption{Examples of connected correlators for the quantum system. (a) The $\langle \hat{S}^z_1\hat{S}^z_{2} \rangle_c$ connected correlator for $\alpha=0$. We can see the connected correlator becomes negative for nearest-neighbour sites when in the mean-field AFM region. The insets (b) and (c) show examples of how the connected correlator varies across sites for the red and blue circle respectively. We see that in both cases, long-range order is lost, but the changes of sign are as expected.}
\vspace{-0cm}
\label{ConnectedCorrelatorsZZ}
\end{figure}

\begin{figure}
	\hspace*{0cm}
	\includegraphics[scale=0.45,clip,angle=0]{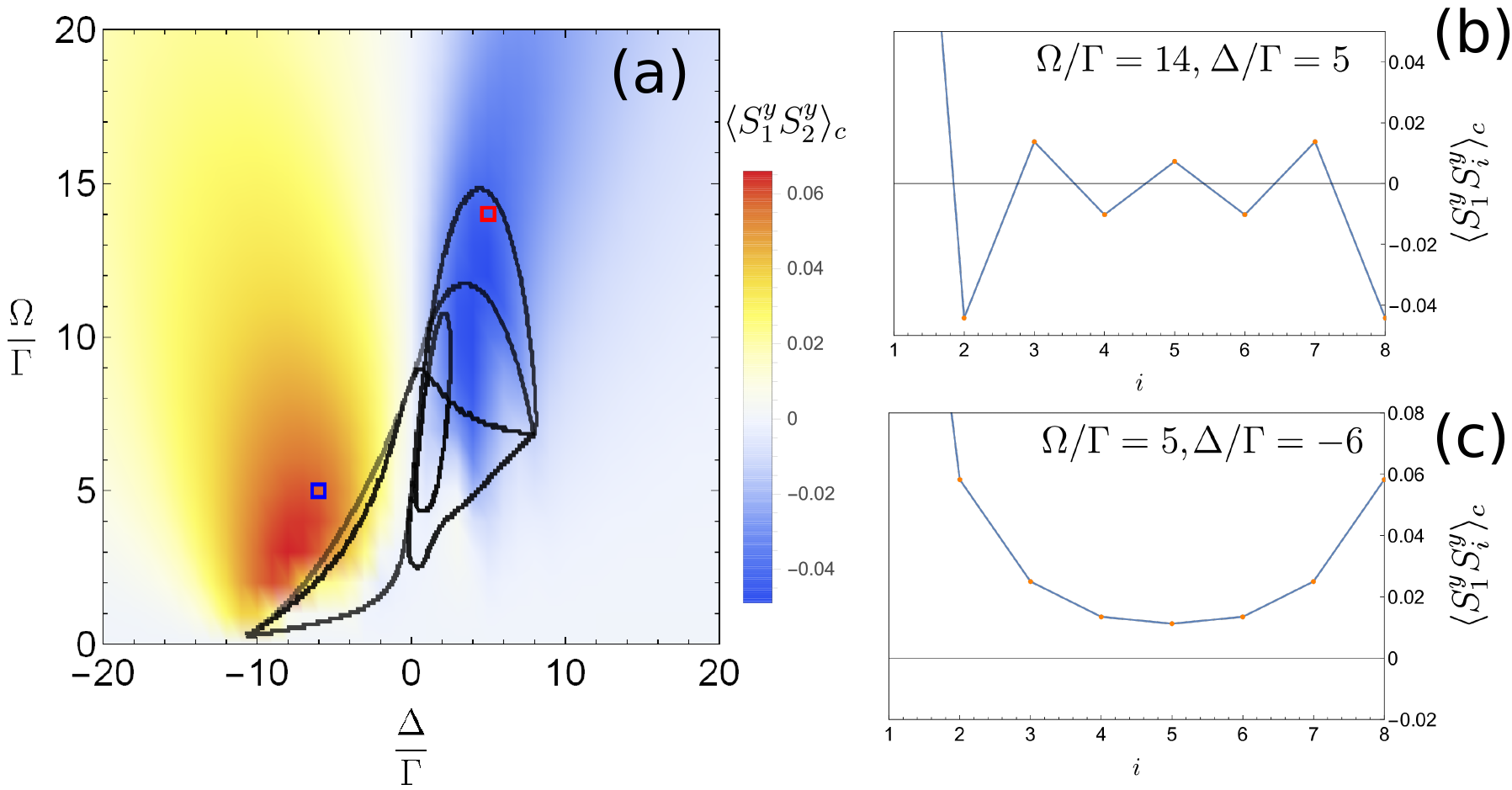}
	\caption{Examples of connected correlators for the quantum system. (a) The $\langle \hat{S}^y_1\hat{S}^y_{2} \rangle_c$ connected correlator for $\alpha=\pi/2$. Again, we can see that the connected correlator changes sign between nearest-neighbours when in the mean-field AFM region. The insets (b) and (c) show examples of how the connected correlator varies across sites for the red and blue squares respectively.}
	\vspace{-0cm}
	\label{ConnectedCorrelatorsXY}
\end{figure}

Although the expectation values only show a single uniform phase, the connected correlators between sites give insight into spatial structure of the fluctuations about the expectation values and possible emergence of non-uniform phases in large systems. In Fig. \ref{ConnectedCorrelatorsZZ} (a), we plot the connected correlator, $\langle \hat{S}^z_i\hat{S}^z_j \rangle_c=\langle \hat{S}^z_i\hat{S}^z_j \rangle-\langle \hat{S}^z_i \rangle\langle \hat{S}^z_j \rangle$, between a site $i$ and its nearest-neighbour for $\alpha=0$. We find that the connected correlator changes sign between nearest-neighbour sites in the region where where AFM solutions exist in the mean-field, but maintains the same sign when in the uniform region. Figure \ref{ConnectedCorrelatorsZZ} (b) and (c) show examples of how the connected correlator varies across the lattice sites for a choice of $\Delta$ and $\Omega$ in the AFM region and in the uniform region, shown by the red and blue circles respectively. We can see how both connected correlators lose long-range order quickly, but maintain an alternating sign in the AFM region, while being persistently positive in the uniform region. We also plot the $\langle \hat{S}^y_i\hat{S}^y_j \rangle_c$ connected correlator for $\alpha=\pi/2$ in Fig. \ref{ConnectedCorrelatorsXY}, finding again a loss of long-range order, but sign changes between nearest-neighbour sites in the region where AFM solutions exist in the mean-field phase diagram. The $\alpha=0$ connected correlator results are similar to those in Ref. \cite{Lee2011a} and the $\alpha=\pi/2$ results similar to those in Ref. \cite{Wilson2016a}. For both Fig. \ref{ConnectedCorrelatorsZZ} and Fig. \ref{ConnectedCorrelatorsXY}, the choice of connected correlator is based on the spin component with the strongest deviation from the spatially uniform phase in the mean-field analysis. 

\begin{figure*}
 		\hspace*{0cm}
 		\includegraphics[scale=0.4,clip,angle=0]{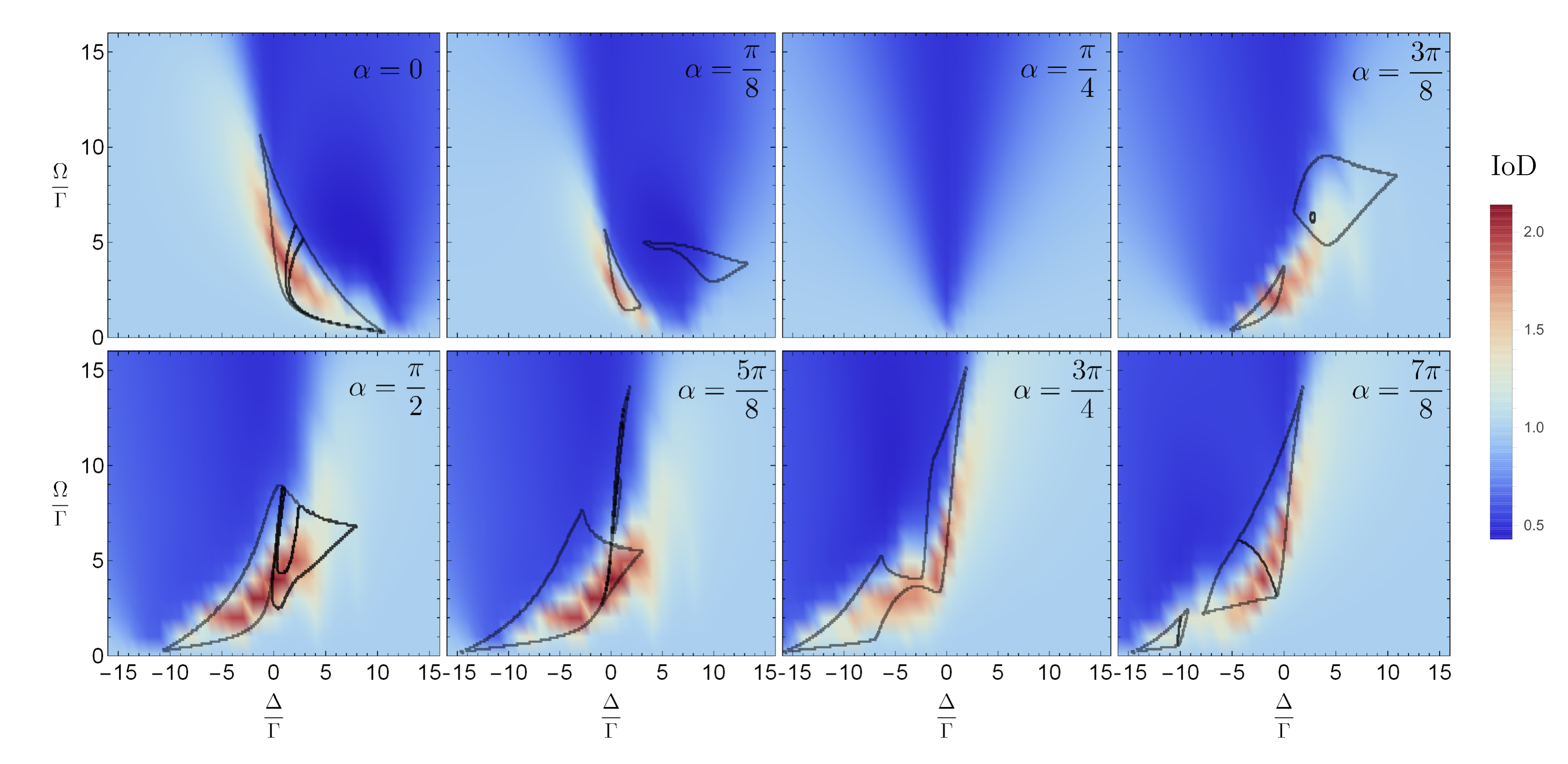}
 		\caption{Index of dispersion (IoD) of the total number of excitations, $\sum_{i}^{N}(\hat{S}^z_i+\hat{\mathbb{1}}/2)$. We see there is an increase in the IoD in regions of bistability, before it drops quickly to $1/2$ when entering the region of the $U_2$ phase. The black contours show the regions of our mean-field phase diagram where bistability occurs.}
 		\vspace{-0cm}
 		\label{FluctuationPlots}
 	\end{figure*}

We find the strongest signature of the mean-field occurs when looking at phase bistability. In regions of bistability, many studies \cite{Wilson2016a,Olmos2014,Lee2012a} have found examples of bimodality in the expectation value distributions when using Quantum Monte Carlo wavefunction methods. This bimodality arises when there are two stable mean-field solutions where, in the mean-field limit, each steady state behaves as if the minimum of a double well potential. Classically, the system will sit in either steady state indefinitely. However, in the full quantum system, quantum fluctuations induce transitions between these steady states, giving a single expectation value, but leaving a double well structure in the expectation value distributions. For the phase diagram in Fig. \ref{phasediagrams}, bistability brings the greatest change in $S^z$ values between the two mean-field phases, and so, for the quantum system, we expect to see bimodality in the total number of excitations, $\sum_{i}^{N}(\hat{S}^z_i+\hat{\mathbb{1}}/2)$.
As first studied in Ref. \cite{Wilson2016a}, a bimodal distribution can be observed by studying the Index of Dispersion (IoD), which is given by
\begin{equation}\label{IoD}
\begin{split}	
\text{IoD}=\frac{\sum_{i,j}^{N}\left(\langle \hat{S}_i^z\hat{S}_j^z\rangle-\langle \hat{S}_i^z\rangle\langle \hat{S}_j^z\rangle\right)}{\sum_{i}^{N}\left(\langle \hat{S}_i^z\rangle+1/2\right)},
\end{split}
\end{equation}
and is a measure of a distribution's variance normalised by its mean. In the limit of zero Rabi drive, when $\langle \hat{S}_i^z \rangle=-1/2$, we have $\text{IoD}=1$, whereas in the limit of high Rabi drive, when $\langle \hat{S}_i^z \rangle \approx 0$, we have $\text{IoD}=1/2$. In between these two limits, the IoD will either decrease or, when there is a bimodality in the distribution, increase above unity. In Fig. \ref{FluctuationPlots}, we plot the IoD as a function of $\Delta$ and $\Omega$.
We can see from Fig. \ref{FluctuationPlots} that the agreement between the fluctuation region and the bistability region is good for all $\alpha$ values, suggesting that the large peak in the IoD is indeed due to the double well structure of the mean-field solution appearing in the quantum steady state. The results for $\alpha=\pi/2$ are similar to those in \cite{Wilson2016a} for the nearest-neighbour XY model, but the results for other $\alpha$ are new.	
	
Another signature of bistability should also be evident in spectral gap of the Liouvillian. The master equation, Eq. \eqref{MasterEq}, can be written as $d\hat{\rho}(t)/dt=\mathcal{L}\hat{\rho}(t)$, where $\mathcal{L}$ is the Liouvillian superoperator that determines the evolution of the system. By expanding the density matrix in eigenvectors of the Liouvillian superoperator, we can write a generic quantum state as 
\begin{equation}\label{superoperator}
\begin{split}	
\hat{\rho}(t)&=\exp(\mathcal{L}t)\hat{\rho}(0)\\
&=\hat{\rho}_{ss}+\sum_{i=1}^{4^N-1}\tr{\hat{L}_i\hat{\rho}(0)}\hat{R}_ie^{\lambda_it},
\end{split}
\end{equation}
where $\hat{L}_i$ and $\hat{R}_i$ are the left and right eigenvectors of $\mathcal{L}$ respectively, with complex eigenvalues, $\lambda_i$. The eigenvalues of the Liouvillian come in complex-conjugate pairs due to the Hermitian nature of the density matrix, with a negative real part corresponding to the decay of the eigenvalues and the imaginary part corresponding to coherences. We order the eigenvalues, $\lambda_{i}$, by the increasing magnitude of their real part. There is always at least one zero eigenvalue of the Liouvillian matrix ($\lambda_0=0$) corresponding to the steady state of the system, which we have extracted from the sum in Eq. \eqref{superoperator} and denoted the eigenvector as $\hat{\rho}_{ss}\equiv\hat{R}_0$. Note that $\hat{\rho}_{ss}$ has unit trace and the remaining eigenvectors, $\hat{R}_i$, are traceless. 

\begin{figure*}
 	\hspace*{0cm}
 	\includegraphics[scale=0.4,clip,angle=0]{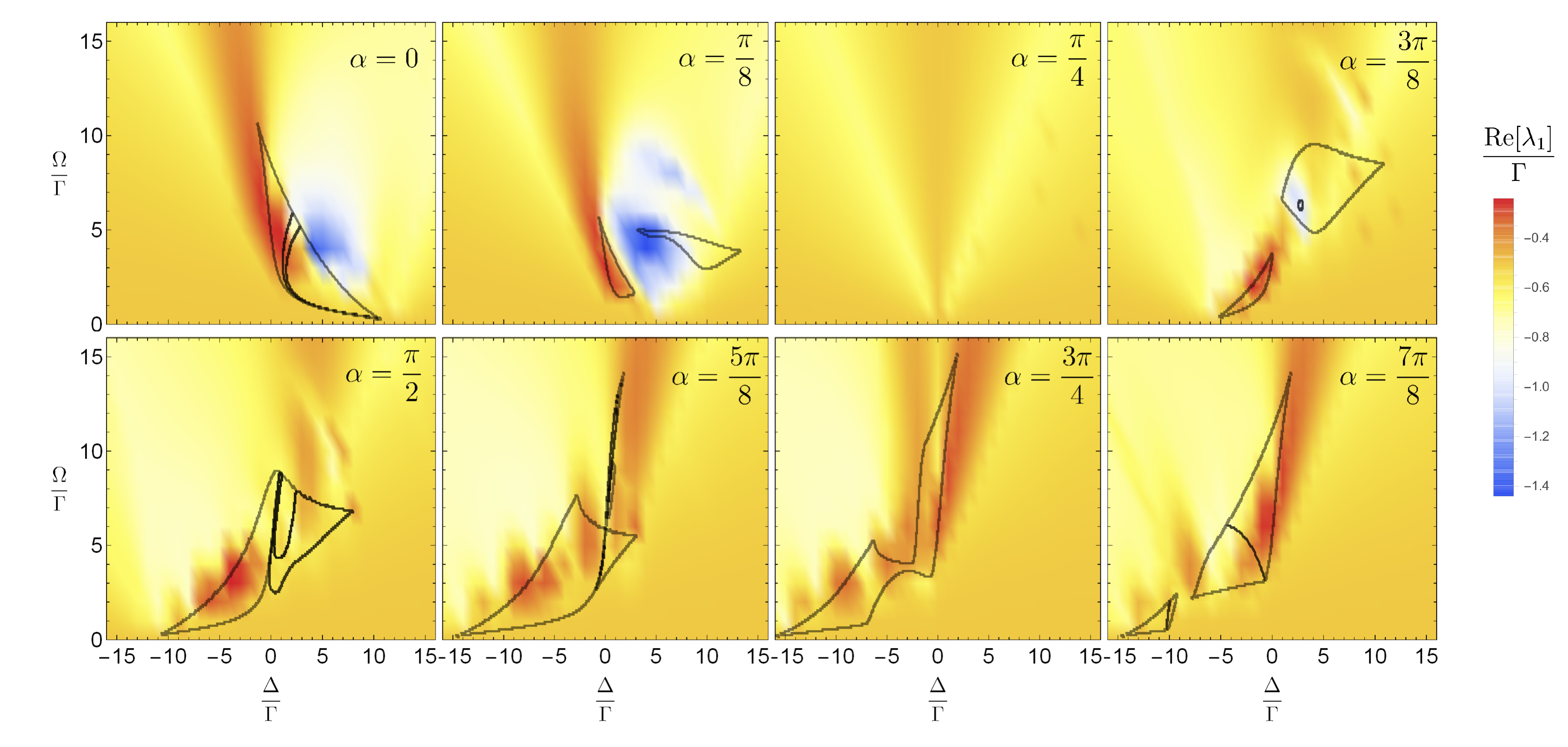}
 	\caption{Real part of the spectral gap of the Liouvillian. We see that this is minimal within the regime of bistability for all $\alpha$. We expect that the spectral gap could close in the limit of large enough quantum systems.}
 	\vspace{-0cm}
 	\label{GapPlots}
 \end{figure*}

In systems with bistability, the spectral gap (i.e. the eigenvalue with smallest real part) should close so the system has two steady states. However, for small quantum systems, it is known that the gap will remain finite \cite{Schirmer2010}. Despite this, bistability may occur in very large quantum systems, and if this is to be the case, then we expect the gap to decrease in the region of bistability even for small systems. In Fig. \ref{GapPlots}, we compute \cite{Navarrete-benllocha} and plot the real part of the spectral gap for the system with $N=6$ spins.
We see there is indeed a reduction in the gap size in the bistability region compared to elsewhere in the phase diagram. We also study the real part of the spectral gap in smaller system sizes ($N=3,4,5$) and find that the real part of the spectral gap decreases with increasing $N$. This may indicate a closing of the gap in larger system sizes, although the gap could also saturate.

As well as examining the real part of the spectral gap, we also analyse the imaginary part. We expect that in regions where there are persistent oscillations, that the real part of the gap should go to zero and the imaginary part should have a non-zero value. However, we have found little sign this was the case and have not shown the data here. This behaviour may be due to the fact that the oscillations come from instabilities of the AFM phases. Our analysis of the real part of the spectral gap showed little decrease in the real part of the spectral gap in regions where there are mean-field bistabilities between the uniform and the AFM phases. Therefore, it appears to be the case that non-uniform phases give weak spectral signatures, which may explain the absence of any spectral signature for oscillations. It would be interesting to examine this further.

\begin{figure*}
\hspace*{0cm}
\includegraphics[scale=0.4,clip,angle=0]{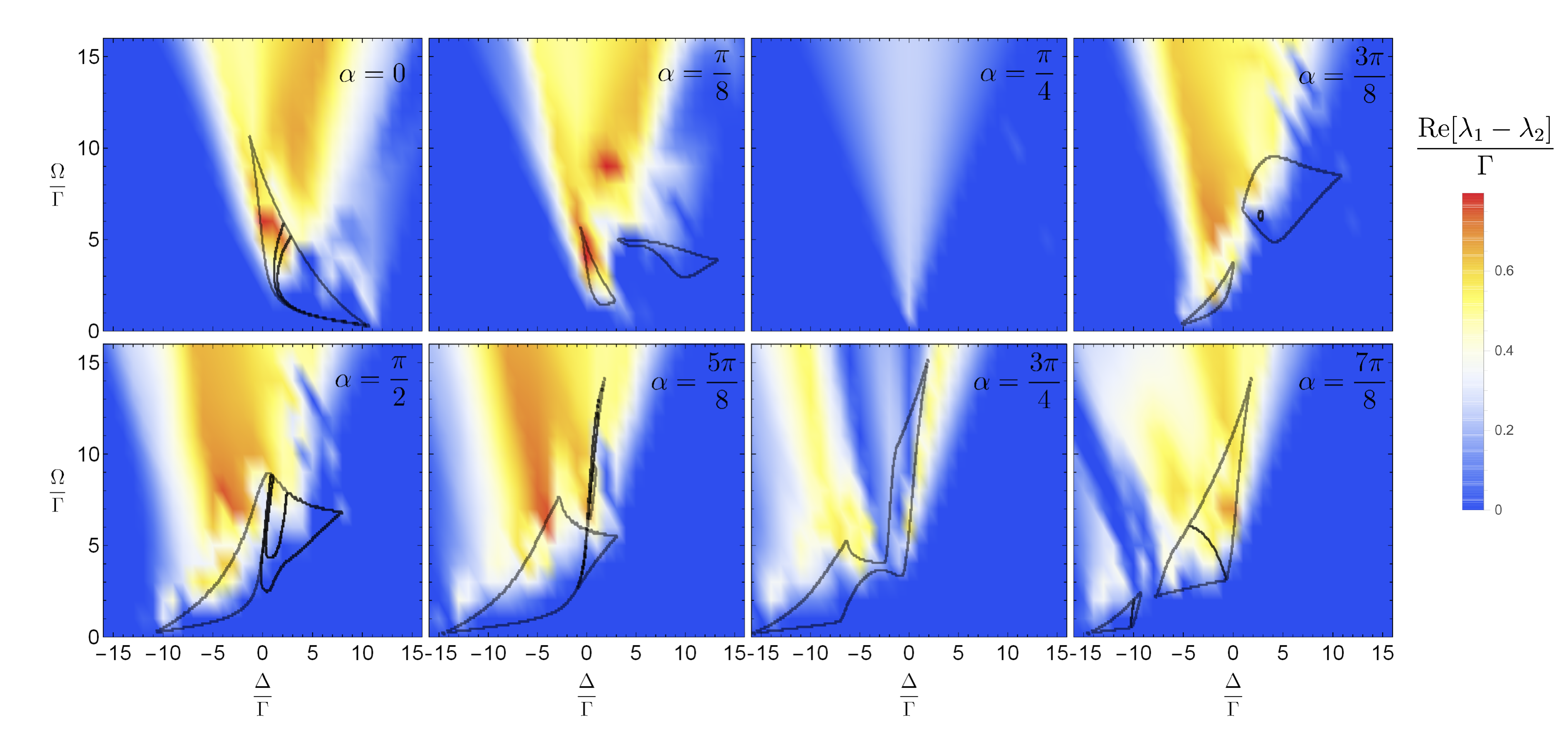}
\caption{Difference between the real part of the spectral gap and the real part of the next eigenvalue in the spectrum. We see for a dominant Ising interaction, there is an increase in separation which indicates the spectral gap could be measured from the decay of the two-time correlator. For an XY dominated interaction, the difference between the real part of the eigenvalues is much smaller and so measuring the gap from the two-time correlator will be harder.}
\vspace{-0cm}
\label{2ndGapPlots}
\end{figure*}

To measure the spectral gap in experiment, one could look at two-time correlators, whose decay depends specifically on the spectral gap in the long-time limit. For two operators, $\hat{A}$ and $\hat{B}$, the two-time correlator is given by \cite{Breuer2007}
\begin{equation}\label{Two-time-correlator}
\begin{split}	
&\langle \hat{A}(t+\tau)\hat{B}(t)\rangle=\Tr\{\hat{A}e^{(\tau+t-t)\mathcal{L}}\hat{B}\hat{\rho}(t)\}\\
&=\Tr\{\hat{B}\hat{\rho}(t)\}\Tr\{\hat{A}\hat{\rho}_{ss}\}+\sum_{i=1}^{4^N-1}e^{\lambda_i\tau}\Tr\{\hat{L}_i\hat{B}\hat{\rho}(t)\}\Tr\{\hat{A}\hat{R}_i\},
\end{split}
\end{equation}
where we have inserted the density matrix expansion given in Eq. \eqref{superoperator}. If we allow $t\rightarrow\infty$ so we study the two-time correlator in the steady state, then Eq. \eqref{Two-time-correlator} simplifies to
\begin{equation}\label{Two-time-correlator-inf}
\begin{split}	
& \lim\limits_{t\rightarrow\infty}\langle \hat{A}(t+\tau)\hat{B}(t)\rangle\\
&=\Tr\{\hat{A}\hat{\rho}_{ss}\}\Tr\{\hat{B}\hat{\rho}_{ss}\}+\sum_{i=1}^{4^N-1}\Tr\{\hat{L}_i\hat{B}\hat{\rho}_{ss}\}\Tr\{\hat{A}\hat{R}_i\}e^{\lambda_i\tau}\\
&\approx\Tr\{\hat{A}\hat{\rho}_{ss}\}\Tr\{\hat{B}\hat{\rho}_{ss}\}+\Tr\{\hat{L}_1\hat{B}\hat{\rho}_{ss}\}\Tr\{\hat{A}\hat{R}_1\}e^{\lambda_1\tau},
\end{split}
\end{equation}
where in the last line we have assumed that $\tau$ is large and that the spectral gap, $\lambda_1$, is well separated from the rest of the Liouvillian spectrum. This shows how the two-time correlator will decay exponentially with a decay time set by the spectral gap. Therefore, if the spectral gap does decrease in a given parameter regime, we should find that any two-time correlator will have a much longer temporal decay than in regions where the spectral gap is large. In Fig. \ref{2ndGapPlots}, we plot the difference in real part of the first and second eigenvalues of the Liouvillian to see if the spectral gap is well separated from the spectral bulk. We find for a dominant Ising interaction, there is a separation inside the bistable regime, so measuring the gap from the connected correlator decay should be possible. For a dominant XY interaction, the gap from the bulk is quite small and therefore measuring the gap from the two-time correlator will be harder.

\section{Discussion}
\label{Discussion}
We have studied the mean-field phase diagram of an driven-dissipative XXZ model with a tunable XY and Ising interaction, finding that the interplay between drive and dissipation leads to four key types of non-equilibrium phases. Specifically, we find uniform phases, AFM phases, SDW phases and OSC phases, as well as bistabilities between these phases, and have studied how these phases change with the tuning of the Ising to XY interaction. Such a system could be readily studied with Rydberg atoms \cite{Glaetzle2015}, trapped ions \cite{Britton2012} or polar molecules \cite{Gorshkov2011,Wall2014}, where dissipation can be controllably induced with optical pumping \cite{Lee2013a}.

Our in-depth study of the full quantum system for a small number of spins shows that the expectation values of the small quantum system agree well with the mean-field uniform phases at strong Rabi drive and detuning, but do not agree as well when the drive and detuning are comparable to the interaction strength. The biggest difference between the mean-field and quantum results is the small quantum system does not exhibit any bistabilities, OSC, AFM or SDW phases. However, analysis of connected correlators shows how fluctuations about these expectation values give signatures of the non-spatially uniform mean-field phases and indicate the possibility of agreement between the mean-field and quantum results for large enough systems. The strongest signatures of the mean-field can be found when looking at mean-field bistability, where we have shown there is good agreement between the regions of mean-field bistability and where the IoD of excitations increases above unity. Finally, we have shown a good agreement between the bistability region and a decrease in the real part of the spectral gap in the Liouvillian, which should be observable by a increase in the decay times of two-time correlators.

As mentioned in earlier sections, many of our results here agree with similar studies with nearest-neighbour interactions \cite{Lee2011a,Wilson2016a}, both at the mean-field and quantum level, indicating that the $1/r^3$ power-law nature of the interactions have little effect on the resultant phases. Given this and the fact that our system is 1D, modelling the system with DMRG \cite{Schollwock2005} to achieve larger system sizes should be possible. It would be interesting to see if the quantum fluctuations grow stronger in larger systems and lead to the breaking of spatial uniformity we have seen in our analysis, resulting in a phase diagram closer to our mean-field results. It would also be interesting to study how the IoD peak grows with system size. When studying smaller system sizes, we find the maximum value of the IoD for each $\alpha$ value grows with system size. If the mean-field is expected to become increasingly correct with increased dimensionality and system size, then the IoD should to drop to zero as the connected correlators in Eq. \eqref{IoD} become zero. However, we do not expect the mean-field to become valid for a 1D system. Looking at how the IoD behaves in higher dimensionalities would also be interesting. Finally, carrying out finite-size scaling of the Liouvillian gap to larger system sizes than studied here may indicate if the gap will eventually close in large enough quantum systems.

\section{Conclusions}
\label{Conclusions}
We have studied the mean-field phase diagram of a driven-dissipative XXZ model with a tunable Ising to XY interaction. We find the emergence of four key types of phase: uniform phases, spin density wave phases, antiferromagnetism and oscillatory phases as well as phase bistabilities. We characterise how the nature of these phases change with the relative strength of the Ising to XY ratio. We then analyse the phases of the corresponding quantum system for small numbers of spins and find that the mean-field results correspond to key features in the quantum phase diagram. We find that peaks in the index of dispersion and decreases in the real part of the Liouvillian gap appear in regimes where the mean-field theory shows bistability. We also find signatures of spatially varying mean-field phases in the connected correlators computed for the full quantum problem. These features persist with changes in the relative strength of the Ising and XY interaction, indicating that they are generic features of spin$-1/2$ systems.

%We have studied the mean-field phase diagram of a driven-dissipative XXZ model with a tunable Ising to XY interaction. We find the emergence of four key types of phase: uniform phases, spin density wave phases, antiferromagnetism and oscillatory phases as well as phase bistabilities. We characterise how the nature of these phases change with the relative strength of the Ising to XY ratio. We then analyse the phases of the corresponding quantum system for small numbers of spins and find that the mean-field results correspond to key features in the quantum phase diagram. We find the strong agreement between peaks in the index of dispersion and a decrease in the real part of the Liouvillian gap, to the regions of mean-field bistability. We also find agreements between connected correlators and spatially varying phases in the mean-field phase diagram. These features persist with changes in the relative strength of the Ising to XY ratio, indicating that they are generic features of spin$-1/2$ systems.

Statement of compliance with EPSRC policy framework on research data: All data accompanying this publication are directly available within the publication.

\section{Acknowledgements}
This work was supported by EPSRC Grant Nos. EP/K030094/1 and EP/P009565/1 and by the Simons Foundation.

\nocite{*}
%\bibliographystyle{apsrev4-1}
%\bibliography{Paper1.5}

\begin{thebibliography}{37}%
	\makeatletter
	\providecommand \@ifxundefined [1]{%
		\@ifx{#1\undefined}
	}%
	\providecommand \@ifnum [1]{%
		\ifnum #1\expandafter \@firstoftwo
		\else \expandafter \@secondoftwo
		\fi
	}%
	\providecommand \@ifx [1]{%
		\ifx #1\expandafter \@firstoftwo
		\else \expandafter \@secondoftwo
		\fi
	}%
	\providecommand \natexlab [1]{#1}%
	\providecommand \enquote  [1]{``#1''}%
	\providecommand \bibnamefont  [1]{#1}%
	\providecommand \bibfnamefont [1]{#1}%
	\providecommand \citenamefont [1]{#1}%
	\providecommand \href@noop [0]{\@secondoftwo}%
	\providecommand \href [0]{\begingroup \@sanitize@url \@href}%
	\providecommand \@href[1]{\@@startlink{#1}\@@href}%
	\providecommand \@@href[1]{\endgroup#1\@@endlink}%
	\providecommand \@sanitize@url [0]{\catcode `\\12\catcode `\$12\catcode
		`\&12\catcode `\#12\catcode `\^12\catcode `\_12\catcode `\%12\relax}%
	\providecommand \@@startlink[1]{}%
	\providecommand \@@endlink[0]{}%
	\providecommand \url  [0]{\begingroup\@sanitize@url \@url }%
	\providecommand \@url [1]{\endgroup\@href {#1}{\urlprefix }}%
	\providecommand \urlprefix  [0]{URL }%
	\providecommand \Eprint [0]{\href }%
	\providecommand \doibase [0]{http://dx.doi.org/}%
	\providecommand \selectlanguage [0]{\@gobble}%
	\providecommand \bibinfo  [0]{\@secondoftwo}%
	\providecommand \bibfield  [0]{\@secondoftwo}%
	\providecommand \translation [1]{[#1]}%
	\providecommand \BibitemOpen [0]{}%
	\providecommand \bibitemStop [0]{}%
	\providecommand \bibitemNoStop [0]{.\EOS\space}%
	\providecommand \EOS [0]{\spacefactor3000\relax}%
	\providecommand \BibitemShut  [1]{\csname bibitem#1\endcsname}%
	\let\auto@bib@innerbib\@empty
	%</preamble>
	\bibitem [{\citenamefont {Chan}\ \emph {et~al.}(2015)\citenamefont {Chan},
		\citenamefont {Lee},\ and\ \citenamefont {Gopalakrishnan}}]{Chan2015b}%
	\BibitemOpen
	\bibfield  {author} {\bibinfo {author} {\bibfnamefont {C.-K.}\ \bibnamefont
			{Chan}}, \bibinfo {author} {\bibfnamefont {T.~E.}\ \bibnamefont {Lee}}, \
		and\ \bibinfo {author} {\bibfnamefont {S.}~\bibnamefont {Gopalakrishnan}},\
	}\href {\doibase 10.1103/PhysRevA.91.051601} {\bibfield  {journal} {\bibinfo
		{journal} {Phys. Rev. A}\ }\textbf {\bibinfo {volume} {91}},\ \bibinfo
	{pages} {051601} (\bibinfo {year} {2015})}\BibitemShut {NoStop}%
\bibitem [{\citenamefont {Lee}\ \emph {et~al.}(2013)\citenamefont {Lee},
	\citenamefont {Gopalakrishnan},\ and\ \citenamefont {Lukin}}]{Lee2013a}%
\BibitemOpen
\bibfield  {author} {\bibinfo {author} {\bibfnamefont {T.~E.}\ \bibnamefont
		{Lee}}, \bibinfo {author} {\bibfnamefont {S.}~\bibnamefont {Gopalakrishnan}},
	\ and\ \bibinfo {author} {\bibfnamefont {M.~D.}\ \bibnamefont {Lukin}},\
}\href {\doibase 10.1103/PhysRevLett.110.257204} {\bibfield  {journal}
{\bibinfo  {journal} {Phys. Rev. Lett.}\ }\textbf {\bibinfo {volume} {110}},\
\bibinfo {pages} {257204} (\bibinfo {year} {2013})}\BibitemShut {NoStop}%
\bibitem [{\citenamefont {Lee}\ \emph {et~al.}(2011)\citenamefont {Lee},
	\citenamefont {H{\"{a}}ffner},\ and\ \citenamefont {Cross}}]{Lee2011a}%
\BibitemOpen
\bibfield  {author} {\bibinfo {author} {\bibfnamefont {T.~E.}\ \bibnamefont
		{Lee}}, \bibinfo {author} {\bibfnamefont {H.}~\bibnamefont {H{\"{a}}ffner}},
	\ and\ \bibinfo {author} {\bibfnamefont {M.~C.}\ \bibnamefont {Cross}},\
}\href {\doibase 10.1103/PhysRevA.84.031402} {\bibfield  {journal} {\bibinfo
	{journal} {Phys. Rev. A}\ }\textbf {\bibinfo {volume} {84}},\ \bibinfo
{pages} {031402} (\bibinfo {year} {2011})}\BibitemShut {NoStop}%
\bibitem [{\citenamefont {Wilson}\ \emph {et~al.}(2016)\citenamefont {Wilson},
	\citenamefont {Mahmud}, \citenamefont {Hu}, \citenamefont {Gorshkov},
	\citenamefont {Hafezi},\ and\ \citenamefont {Foss-Feig}}]{Wilson2016a}%
\BibitemOpen
\bibfield  {author} {\bibinfo {author} {\bibfnamefont {R.~M.}\ \bibnamefont
		{Wilson}}, \bibinfo {author} {\bibfnamefont {K.~W.}\ \bibnamefont {Mahmud}},
	\bibinfo {author} {\bibfnamefont {A.}~\bibnamefont {Hu}}, \bibinfo {author}
	{\bibfnamefont {A.~V.}\ \bibnamefont {Gorshkov}}, \bibinfo {author}
	{\bibfnamefont {M.}~\bibnamefont {Hafezi}}, \ and\ \bibinfo {author}
	{\bibfnamefont {M.}~\bibnamefont {Foss-Feig}},\ }\href {\doibase
	10.1103/PhysRevA.94.033801} {\bibfield  {journal} {\bibinfo  {journal} {Phys.
			Rev. A}\ }\textbf {\bibinfo {volume} {94}},\ \bibinfo {pages} {033801}
	(\bibinfo {year} {2016})}\BibitemShut {NoStop}%
\bibitem [{\citenamefont {Qian}\ \emph {et~al.}(2015)\citenamefont {Qian},
	\citenamefont {Zhang}, \citenamefont {Zhai},\ and\ \citenamefont
	{Zhang}}]{Qian2015}%
\BibitemOpen
\bibfield  {author} {\bibinfo {author} {\bibfnamefont {J.}~\bibnamefont
		{Qian}}, \bibinfo {author} {\bibfnamefont {L.}~\bibnamefont {Zhang}},
	\bibinfo {author} {\bibfnamefont {J.}~\bibnamefont {Zhai}}, \ and\ \bibinfo
	{author} {\bibfnamefont {W.}~\bibnamefont {Zhang}},\ }\href {\doibase
	10.1103/PhysRevA.92.063407} {\bibfield  {journal} {\bibinfo  {journal} {Phys.
			Rev. A}\ }\textbf {\bibinfo {volume} {92}},\ \bibinfo {pages} {063407}
	(\bibinfo {year} {2015})}\BibitemShut {NoStop}%
\bibitem [{\citenamefont {Parmee}\ and\ \citenamefont
	{Cooper}(2018)}]{Parmee2018}%
\BibitemOpen
\bibfield  {author} {\bibinfo {author} {\bibfnamefont {C.~D.}\ \bibnamefont
		{Parmee}}\ and\ \bibinfo {author} {\bibfnamefont {N.~R.}\ \bibnamefont
		{Cooper}},\ }\href {\doibase 10.1103/PhysRevA.97.053616} {\bibfield
	{journal} {\bibinfo  {journal} {Phys. Rev. A}\ }\textbf {\bibinfo {volume}
		{97}},\ \bibinfo {pages} {053616} (\bibinfo {year} {2018})}\BibitemShut
{NoStop}%
\bibitem [{\citenamefont {Qian}\ \emph {et~al.}(2012)\citenamefont {Qian},
	\citenamefont {Dong}, \citenamefont {Zhou},\ and\ \citenamefont
	{Zhang}}]{Qian2012a}%
\BibitemOpen
\bibfield  {author} {\bibinfo {author} {\bibfnamefont {J.}~\bibnamefont
		{Qian}}, \bibinfo {author} {\bibfnamefont {G.}~\bibnamefont {Dong}}, \bibinfo
	{author} {\bibfnamefont {L.}~\bibnamefont {Zhou}}, \ and\ \bibinfo {author}
	{\bibfnamefont {W.}~\bibnamefont {Zhang}},\ }\href {\doibase
	10.1103/PhysRevA.85.065401} {\bibfield  {journal} {\bibinfo  {journal} {Phys.
			Rev. A}\ }\textbf {\bibinfo {volume} {85}},\ \bibinfo {pages} {065401}
	(\bibinfo {year} {2012})}\BibitemShut {NoStop}%
\bibitem [{\citenamefont {Carr}\ \emph {et~al.}(2013)\citenamefont {Carr},
	\citenamefont {Ritter}, \citenamefont {Wade}, \citenamefont {Adams},\ and\
	\citenamefont {Weatherill}}]{Carr2013}%
\BibitemOpen
\bibfield  {author} {\bibinfo {author} {\bibfnamefont {C.}~\bibnamefont
		{Carr}}, \bibinfo {author} {\bibfnamefont {R.}~\bibnamefont {Ritter}},
	\bibinfo {author} {\bibfnamefont {C.~G.}\ \bibnamefont {Wade}}, \bibinfo
	{author} {\bibfnamefont {C.~S.}\ \bibnamefont {Adams}}, \ and\ \bibinfo
	{author} {\bibfnamefont {K.~J.}\ \bibnamefont {Weatherill}},\ }\href
{\doibase 10.1103/PhysRevLett.111.113901} {\bibfield  {journal} {\bibinfo
		{journal} {Phys. Rev. Lett.}\ }\textbf {\bibinfo {volume} {111}},\ \bibinfo
	{pages} {113901} (\bibinfo {year} {2013})}\BibitemShut {NoStop}%
\bibitem [{\citenamefont {{\v{S}}ibali{\'{c}}}\ \emph
	{et~al.}(2016)\citenamefont {{\v{S}}ibali{\'{c}}}, \citenamefont {Wade},
	\citenamefont {Adams}, \citenamefont {Weatherill},\ and\ \citenamefont
	{Pohl}}]{Sibalic2016}%
\BibitemOpen
\bibfield  {author} {\bibinfo {author} {\bibfnamefont {N.}~\bibnamefont
		{{\v{S}}ibali{\'{c}}}}, \bibinfo {author} {\bibfnamefont {C.~G.}\
		\bibnamefont {Wade}}, \bibinfo {author} {\bibfnamefont {C.~S.}\ \bibnamefont
		{Adams}}, \bibinfo {author} {\bibfnamefont {K.~J.}\ \bibnamefont
		{Weatherill}}, \ and\ \bibinfo {author} {\bibfnamefont {T.}~\bibnamefont
		{Pohl}},\ }\href {\doibase 10.1103/PhysRevA.94.011401} {\bibfield  {journal}
	{\bibinfo  {journal} {Phys. Rev. A}\ }\textbf {\bibinfo {volume} {94}},\
	\bibinfo {pages} {011401} (\bibinfo {year} {2016})}\BibitemShut {NoStop}%
\bibitem [{\citenamefont {Olmos}\ \emph {et~al.}(2014)\citenamefont {Olmos},
	\citenamefont {Yu},\ and\ \citenamefont {Lesanovsky}}]{Olmos2014}%
\BibitemOpen
\bibfield  {author} {\bibinfo {author} {\bibfnamefont {B.}~\bibnamefont
		{Olmos}}, \bibinfo {author} {\bibfnamefont {D.}~\bibnamefont {Yu}}, \ and\
	\bibinfo {author} {\bibfnamefont {I.}~\bibnamefont {Lesanovsky}},\ }\href
{\doibase 10.1103/PhysRevA.89.023616} {\bibfield  {journal} {\bibinfo
		{journal} {Phys. Rev. A}\ }\textbf {\bibinfo {volume} {89}},\ \bibinfo
	{pages} {023616} (\bibinfo {year} {2014})}\BibitemShut {NoStop}%
\bibitem [{\citenamefont {Rota}\ \emph {et~al.}(2018)\citenamefont {Rota},
	\citenamefont {Minganti}, \citenamefont {Biella},\ and\ \citenamefont
	{Ciuti}}]{Rota2018}%
\BibitemOpen
\bibfield  {author} {\bibinfo {author} {\bibfnamefont {R.}~\bibnamefont
		{Rota}}, \bibinfo {author} {\bibfnamefont {F.}~\bibnamefont {Minganti}},
	\bibinfo {author} {\bibfnamefont {A.}~\bibnamefont {Biella}}, \ and\ \bibinfo
	{author} {\bibfnamefont {C.}~\bibnamefont {Ciuti}},\ }\href {\doibase
	10.1088/1367-2630/aab703} {\bibfield  {journal} {\bibinfo  {journal} {New J.
			Phys.}\ }\textbf {\bibinfo {volume} {20}},\ \bibinfo {pages} {045003}
	(\bibinfo {year} {2018})}\BibitemShut {NoStop}%
\bibitem [{\citenamefont {Lee}\ \emph {et~al.}(2012)\citenamefont {Lee},
	\citenamefont {H{\"{a}}ffner},\ and\ \citenamefont {Cross}}]{Lee2012a}%
\BibitemOpen
\bibfield  {author} {\bibinfo {author} {\bibfnamefont {T.~E.}\ \bibnamefont
		{Lee}}, \bibinfo {author} {\bibfnamefont {H.}~\bibnamefont {H{\"{a}}ffner}},
	\ and\ \bibinfo {author} {\bibfnamefont {M.~C.}\ \bibnamefont {Cross}},\
}\href {\doibase 10.1103/PhysRevLett.108.023602} {\bibfield  {journal}
{\bibinfo  {journal} {Phys. Rev. Lett.}\ }\textbf {\bibinfo {volume} {108}},\
\bibinfo {pages} {023602} (\bibinfo {year} {2012})}\BibitemShut {NoStop}%
\bibitem [{\citenamefont {Ates}\ \emph {et~al.}(2012)\citenamefont {Ates},
	\citenamefont {Olmos}, \citenamefont {Garrahan},\ and\ \citenamefont
	{Lesanovsky}}]{Ates2012a}%
\BibitemOpen
\bibfield  {author} {\bibinfo {author} {\bibfnamefont {C.}~\bibnamefont
		{Ates}}, \bibinfo {author} {\bibfnamefont {B.}~\bibnamefont {Olmos}},
	\bibinfo {author} {\bibfnamefont {J.~P.}\ \bibnamefont {Garrahan}}, \ and\
	\bibinfo {author} {\bibfnamefont {I.}~\bibnamefont {Lesanovsky}},\ }\href
{\doibase 10.1103/PhysRevA.85.043620} {\bibfield  {journal} {\bibinfo
		{journal} {Phys. Rev. A}\ }\textbf {\bibinfo {volume} {85}},\ \bibinfo
	{pages} {043620} (\bibinfo {year} {2012})}\BibitemShut {NoStop}%
\bibitem [{\citenamefont {Maghrebi}\ and\ \citenamefont
	{Gorshkov}(2016)}]{Maghrebi2016}%
\BibitemOpen
\bibfield  {author} {\bibinfo {author} {\bibfnamefont {M.~F.}\ \bibnamefont
		{Maghrebi}}\ and\ \bibinfo {author} {\bibfnamefont {A.~V.}\ \bibnamefont
		{Gorshkov}},\ }\href {\doibase 10.1103/PhysRevB.93.014307} {\bibfield
	{journal} {\bibinfo  {journal} {Phys. Rev. B}\ }\textbf {\bibinfo {volume}
		{93}},\ \bibinfo {pages} {014307} (\bibinfo {year} {2016})}\BibitemShut
{NoStop}%
\bibitem [{\citenamefont {Jin}\ \emph {et~al.}(2016)\citenamefont {Jin},
	\citenamefont {Biella}, \citenamefont {Viyuela}, \citenamefont {Mazza},
	\citenamefont {Keeling}, \citenamefont {Fazio},\ and\ \citenamefont
	{Rossini}}]{Jin2016}%
\BibitemOpen
\bibfield  {author} {\bibinfo {author} {\bibfnamefont {J.}~\bibnamefont
		{Jin}}, \bibinfo {author} {\bibfnamefont {A.}~\bibnamefont {Biella}},
	\bibinfo {author} {\bibfnamefont {O.}~\bibnamefont {Viyuela}}, \bibinfo
	{author} {\bibfnamefont {L.}~\bibnamefont {Mazza}}, \bibinfo {author}
	{\bibfnamefont {J.}~\bibnamefont {Keeling}}, \bibinfo {author} {\bibfnamefont
		{R.}~\bibnamefont {Fazio}}, \ and\ \bibinfo {author} {\bibfnamefont
		{D.}~\bibnamefont {Rossini}},\ }\href {\doibase 10.1103/PhysRevX.6.031011}
{\bibfield  {journal} {\bibinfo  {journal} {Phys. Rev. X}\ }\textbf {\bibinfo
		{volume} {6}},\ \bibinfo {pages} {031011} (\bibinfo {year}
	{2016})}\BibitemShut {NoStop}%
\bibitem [{\citenamefont {Jin}\ \emph {et~al.}(2018)\citenamefont {Jin},
	\citenamefont {Biella}, \citenamefont {Viyuela}, \citenamefont {Ciuti},
	\citenamefont {Fazio},\ and\ \citenamefont {Rossini}}]{Jin2018}%
\BibitemOpen
\bibfield  {author} {\bibinfo {author} {\bibfnamefont {J.}~\bibnamefont
		{Jin}}, \bibinfo {author} {\bibfnamefont {A.}~\bibnamefont {Biella}},
	\bibinfo {author} {\bibfnamefont {O.}~\bibnamefont {Viyuela}}, \bibinfo
	{author} {\bibfnamefont {C.}~\bibnamefont {Ciuti}}, \bibinfo {author}
	{\bibfnamefont {R.}~\bibnamefont {Fazio}}, \ and\ \bibinfo {author}
	{\bibfnamefont {D.}~\bibnamefont {Rossini}},\ }\href {\doibase
	10.1103/PhysRevB.98.241108} {\bibfield  {journal} {\bibinfo  {journal} {Phys.
			Rev. B}\ }\textbf {\bibinfo {volume} {98}},\ \bibinfo {pages} {241108}
	(\bibinfo {year} {2018})}\BibitemShut {NoStop}%
\bibitem [{\citenamefont {Owen}\ \emph {et~al.}(2018)\citenamefont {Owen},
	\citenamefont {Jin}, \citenamefont {Rossini}, \citenamefont {Fazio},\ and\
	\citenamefont {Hartmann}}]{Owen2018}%
\BibitemOpen
\bibfield  {author} {\bibinfo {author} {\bibfnamefont {E.~T.}\ \bibnamefont
		{Owen}}, \bibinfo {author} {\bibfnamefont {J.}~\bibnamefont {Jin}}, \bibinfo
	{author} {\bibfnamefont {D.}~\bibnamefont {Rossini}}, \bibinfo {author}
	{\bibfnamefont {R.}~\bibnamefont {Fazio}}, \ and\ \bibinfo {author}
	{\bibfnamefont {M.~J.}\ \bibnamefont {Hartmann}},\ }\href {\doibase
	10.1088/1367-2630/aab7d3} {\bibfield  {journal} {\bibinfo  {journal} {New J.
			Phys.}\ }\textbf {\bibinfo {volume} {20}},\ \bibinfo {pages} {045004}
	(\bibinfo {year} {2018})}\BibitemShut {NoStop}%
\bibitem [{\citenamefont {Biella}\ \emph {et~al.}(2018)\citenamefont {Biella},
	\citenamefont {Jin}, \citenamefont {Viyuela}, \citenamefont {Ciuti},
	\citenamefont {Fazio},\ and\ \citenamefont {Rossini}}]{Biella2018}%
\BibitemOpen
\bibfield  {author} {\bibinfo {author} {\bibfnamefont {A.}~\bibnamefont
		{Biella}}, \bibinfo {author} {\bibfnamefont {J.}~\bibnamefont {Jin}},
	\bibinfo {author} {\bibfnamefont {O.}~\bibnamefont {Viyuela}}, \bibinfo
	{author} {\bibfnamefont {C.}~\bibnamefont {Ciuti}}, \bibinfo {author}
	{\bibfnamefont {R.}~\bibnamefont {Fazio}}, \ and\ \bibinfo {author}
	{\bibfnamefont {D.}~\bibnamefont {Rossini}},\ }\href {\doibase
	10.1103/PhysRevB.97.035103} {\bibfield  {journal} {\bibinfo  {journal} {Phys.
			Rev. B}\ }\textbf {\bibinfo {volume} {97}},\ \bibinfo {pages} {035103}
	(\bibinfo {year} {2018})}\BibitemShut {NoStop}%
\bibitem [{\citenamefont {Huybrechts}\ and\ \citenamefont
	{Wouters}(2019)}]{Huybrechts2019}%
\BibitemOpen
\bibfield  {author} {\bibinfo {author} {\bibfnamefont {D.}~\bibnamefont
		{Huybrechts}}\ and\ \bibinfo {author} {\bibfnamefont {M.}~\bibnamefont
		{Wouters}},\ }\href {\doibase 10.1103/PhysRevA.99.043841} {\bibfield
	{journal} {\bibinfo  {journal} {Phys. Rev. A}\ }\textbf {\bibinfo {volume}
		{99}},\ \bibinfo {pages} {043841} (\bibinfo {year} {2019})}\BibitemShut
{NoStop}%
\bibitem [{\citenamefont {Rota}\ \emph {et~al.}(2017)\citenamefont {Rota},
	\citenamefont {Storme}, \citenamefont {Bartolo}, \citenamefont {Fazio},\ and\
	\citenamefont {Ciuti}}]{Rota2017}%
\BibitemOpen
\bibfield  {author} {\bibinfo {author} {\bibfnamefont {R.}~\bibnamefont
		{Rota}}, \bibinfo {author} {\bibfnamefont {F.}~\bibnamefont {Storme}},
	\bibinfo {author} {\bibfnamefont {N.}~\bibnamefont {Bartolo}}, \bibinfo
	{author} {\bibfnamefont {R.}~\bibnamefont {Fazio}}, \ and\ \bibinfo {author}
	{\bibfnamefont {C.}~\bibnamefont {Ciuti}},\ }\href {\doibase
	10.1103/PhysRevB.95.134431} {\bibfield  {journal} {\bibinfo  {journal} {Phys.
			Rev. B}\ }\textbf {\bibinfo {volume} {95}},\ \bibinfo {pages} {134431}
	(\bibinfo {year} {2017})}\BibitemShut {NoStop}%
\bibitem [{\citenamefont {Weimer}(2015)}]{Weimer2015}%
\BibitemOpen
\bibfield  {author} {\bibinfo {author} {\bibfnamefont {H.}~\bibnamefont
		{Weimer}},\ }\href {\doibase 10.1103/PhysRevLett.114.040402} {\bibfield
	{journal} {\bibinfo  {journal} {Phys. Rev. Lett.}\ }\textbf {\bibinfo
		{volume} {114}},\ \bibinfo {pages} {040402} (\bibinfo {year}
	{2015})}\BibitemShut {NoStop}%
\bibitem [{\citenamefont {Cui}\ \emph {et~al.}(2015)\citenamefont {Cui},
	\citenamefont {Cirac},\ and\ \citenamefont {Ba{\~{n}}uls}}]{Cui2015}%
\BibitemOpen
\bibfield  {author} {\bibinfo {author} {\bibfnamefont {J.}~\bibnamefont
		{Cui}}, \bibinfo {author} {\bibfnamefont {J.~I.}\ \bibnamefont {Cirac}}, \
	and\ \bibinfo {author} {\bibfnamefont {M.~C.}\ \bibnamefont {Ba{\~{n}}uls}},\
}\href {\doibase 10.1103/PhysRevLett.114.220601} {\bibfield  {journal}
{\bibinfo  {journal} {Phys. Rev. Lett.}\ }\textbf {\bibinfo {volume} {114}},\
\bibinfo {pages} {220601} (\bibinfo {year} {2015})}\BibitemShut {NoStop}%
\bibitem [{\citenamefont {Mascarenhas}\ \emph {et~al.}(2015)\citenamefont
	{Mascarenhas}, \citenamefont {Flayac},\ and\ \citenamefont
	{Savona}}]{Mascarenhas2015}%
\BibitemOpen
\bibfield  {author} {\bibinfo {author} {\bibfnamefont {E.}~\bibnamefont
		{Mascarenhas}}, \bibinfo {author} {\bibfnamefont {H.}~\bibnamefont {Flayac}},
	\ and\ \bibinfo {author} {\bibfnamefont {V.}~\bibnamefont {Savona}},\ }\href
{\doibase 10.1103/PhysRevA.92.022116} {\bibfield  {journal} {\bibinfo
		{journal} {Phys. Rev. A}\ }\textbf {\bibinfo {volume} {92}},\ \bibinfo
	{pages} {022116} (\bibinfo {year} {2015})}\BibitemShut {NoStop}%
\bibitem [{\citenamefont {Mendoza-Arenas}\ \emph {et~al.}(2016)\citenamefont
	{Mendoza-Arenas}, \citenamefont {Clark}, \citenamefont {Felicetti},
	\citenamefont {Romero}, \citenamefont {Solano}, \citenamefont {Angelakis},\
	and\ \citenamefont {Jaksch}}]{Clark2016}%
\BibitemOpen
\bibfield  {author} {\bibinfo {author} {\bibfnamefont {J.~J.}\ \bibnamefont
		{Mendoza-Arenas}}, \bibinfo {author} {\bibfnamefont {S.~R.}\ \bibnamefont
		{Clark}}, \bibinfo {author} {\bibfnamefont {S.}~\bibnamefont {Felicetti}},
	\bibinfo {author} {\bibfnamefont {G.}~\bibnamefont {Romero}}, \bibinfo
	{author} {\bibfnamefont {E.}~\bibnamefont {Solano}}, \bibinfo {author}
	{\bibfnamefont {D.~G.}\ \bibnamefont {Angelakis}}, \ and\ \bibinfo {author}
	{\bibfnamefont {D.}~\bibnamefont {Jaksch}},\ }\href {\doibase
	10.1103/PhysRevA.93.023821} {\bibfield  {journal} {\bibinfo  {journal} {Phys.
			Rev. A}\ }\textbf {\bibinfo {volume} {93}},\ \bibinfo {pages} {023821}
	(\bibinfo {year} {2016})}\BibitemShut {NoStop}%
\bibitem [{\citenamefont {Joshi}\ \emph {et~al.}(2013)\citenamefont {Joshi},
	\citenamefont {Nissen},\ and\ \citenamefont {Keeling}}]{Joshi2013a}%
\BibitemOpen
\bibfield  {author} {\bibinfo {author} {\bibfnamefont {C.}~\bibnamefont
		{Joshi}}, \bibinfo {author} {\bibfnamefont {F.}~\bibnamefont {Nissen}}, \
	and\ \bibinfo {author} {\bibfnamefont {J.}~\bibnamefont {Keeling}},\ }\href
{\doibase 10.1103/PhysRevA.88.063835} {\bibfield  {journal} {\bibinfo
		{journal} {Phys. Rev. A}\ }\textbf {\bibinfo {volume} {88}},\ \bibinfo
	{pages} {063835} (\bibinfo {year} {2013})}\BibitemShut {NoStop}%
\bibitem [{\citenamefont {Kshetrimayum}\ \emph {et~al.}(2017)\citenamefont
	{Kshetrimayum}, \citenamefont {Weimer},\ and\ \citenamefont
	{Or{\'{u}}s}}]{Kshetrimayum2017}%
\BibitemOpen
\bibfield  {author} {\bibinfo {author} {\bibfnamefont {A.}~\bibnamefont
		{Kshetrimayum}}, \bibinfo {author} {\bibfnamefont {H.}~\bibnamefont
		{Weimer}}, \ and\ \bibinfo {author} {\bibfnamefont {R.}~\bibnamefont
		{Or{\'{u}}s}},\ }\href {\doibase 10.1038/s41467-017-01511-6} {\bibfield
	{journal} {\bibinfo  {journal} {Nat. Commun.}\ }\textbf {\bibinfo {volume}
		{8}},\ \bibinfo {pages} {1291} (\bibinfo {year} {2017})}\BibitemShut
{NoStop}%
\bibitem [{\citenamefont {H{\"{o}}ning}\ \emph {et~al.}(2013)\citenamefont
	{H{\"{o}}ning}, \citenamefont {Muth}, \citenamefont {Petrosyan},\ and\
	\citenamefont {Fleischhauer}}]{Honing2013}%
\BibitemOpen
\bibfield  {author} {\bibinfo {author} {\bibfnamefont {M.}~\bibnamefont
		{H{\"{o}}ning}}, \bibinfo {author} {\bibfnamefont {D.}~\bibnamefont {Muth}},
	\bibinfo {author} {\bibfnamefont {D.}~\bibnamefont {Petrosyan}}, \ and\
	\bibinfo {author} {\bibfnamefont {M.}~\bibnamefont {Fleischhauer}},\ }\href
{\doibase 10.1103/PhysRevA.87.023401} {\bibfield  {journal} {\bibinfo
		{journal} {Phys. Rev. A}\ }\textbf {\bibinfo {volume} {87}},\ \bibinfo
	{pages} {023401} (\bibinfo {year} {2013})}\BibitemShut {NoStop}%
\bibitem [{\citenamefont {Hu}\ \emph {et~al.}(2013)\citenamefont {Hu},
	\citenamefont {Lee},\ and\ \citenamefont {Clark}}]{Hu2013}%
\BibitemOpen
\bibfield  {author} {\bibinfo {author} {\bibfnamefont {A.}~\bibnamefont
		{Hu}}, \bibinfo {author} {\bibfnamefont {T.~E.}\ \bibnamefont {Lee}}, \ and\
	\bibinfo {author} {\bibfnamefont {C.~W.}\ \bibnamefont {Clark}},\ }\href
{\doibase 10.1103/PhysRevA.88.053627} {\bibfield  {journal} {\bibinfo
		{journal} {Phys. Rev. A}\ }\textbf {\bibinfo {volume} {88}},\ \bibinfo
	{pages} {053627} (\bibinfo {year} {2013})}\BibitemShut {NoStop}%
\bibitem [{\citenamefont {Gorshkov}\ \emph {et~al.}(2011)\citenamefont
	{Gorshkov}, \citenamefont {Manmana}, \citenamefont {Chen}, \citenamefont
	{Demler}, \citenamefont {Lukin},\ and\ \citenamefont {Rey}}]{Gorshkov2011}%
\BibitemOpen
\bibfield  {author} {\bibinfo {author} {\bibfnamefont {A.~V.}\ \bibnamefont
		{Gorshkov}}, \bibinfo {author} {\bibfnamefont {S.~R.}\ \bibnamefont
		{Manmana}}, \bibinfo {author} {\bibfnamefont {G.}~\bibnamefont {Chen}},
	\bibinfo {author} {\bibfnamefont {E.}~\bibnamefont {Demler}}, \bibinfo
	{author} {\bibfnamefont {M.~D.}\ \bibnamefont {Lukin}}, \ and\ \bibinfo
	{author} {\bibfnamefont {A.~M.}\ \bibnamefont {Rey}},\ }\href {\doibase
	10.1103/PhysRevA.84.033619} {\bibfield  {journal} {\bibinfo  {journal} {Phys.
			Rev. A}\ }\textbf {\bibinfo {volume} {84}},\ \bibinfo {pages} {033619}
	(\bibinfo {year} {2011})}\BibitemShut {NoStop}%
\bibitem [{\citenamefont {Peter}(2015)}]{Peter}%
\BibitemOpen
\bibfield  {author} {\bibinfo {author} {\bibfnamefont {D.}~\bibnamefont
		{Peter}},\ }\emph {\bibinfo {title} {{Quantum states with topological
			properties via dipolar interactions}}},\ \href@noop {} {Ph.D. thesis},\
\bibinfo  {school} {University of Stuttgart} (\bibinfo {year}
{2015})\BibitemShut {NoStop}%
\bibitem [{\citenamefont {Schirmer}\ and\ \citenamefont
	{Wang}(2010)}]{Schirmer2010}%
\BibitemOpen
\bibfield  {author} {\bibinfo {author} {\bibfnamefont {S.~G.}\ \bibnamefont
		{Schirmer}}\ and\ \bibinfo {author} {\bibfnamefont {X.}~\bibnamefont
		{Wang}},\ }\href {\doibase 10.1103/PhysRevA.81.062306} {\bibfield  {journal}
	{\bibinfo  {journal} {Phys. Rev. A}\ }\textbf {\bibinfo {volume} {81}},\
	\bibinfo {pages} {062306} (\bibinfo {year} {2010})}\BibitemShut {NoStop}%
\bibitem [{\citenamefont {Navarrete-Benlloch}(2015)}]{Navarrete-benllocha}%
\BibitemOpen
\bibfield  {author} {\bibinfo {author} {\bibfnamefont {C.}~\bibnamefont
		{Navarrete-Benlloch}},\ }\href {http://arxiv.org/abs/1504.05266} {\
	(\bibinfo {year} {2015})},\ \Eprint {http://arxiv.org/abs/1504.05266}
{arXiv:1504.05266} \BibitemShut {NoStop}%
\bibitem [{\citenamefont {Breuer}\ and\ \citenamefont
	{Petruccione}(2007)}]{Breuer2007}%
\BibitemOpen
\bibfield  {author} {\bibinfo {author} {\bibfnamefont {H.-P.}\ \bibnamefont
		{Breuer}}\ and\ \bibinfo {author} {\bibfnamefont {F.}~\bibnamefont
		{Petruccione}},\ }\href {\doibase 10.1093/acprof:oso/9780199213900.001.0001}
{\emph {\bibinfo {title} {{The Theory of Open Quantum Systems}}}}\ (\bibinfo
{publisher} {Oxford University Press},\ \bibinfo {year} {2007})\BibitemShut
{NoStop}%
\bibitem [{\citenamefont {Glaetzle}\ \emph {et~al.}(2015)\citenamefont
	{Glaetzle}, \citenamefont {Dalmonte}, \citenamefont {Nath}, \citenamefont
	{Gross}, \citenamefont {Bloch},\ and\ \citenamefont {Zoller}}]{Glaetzle2015}%
\BibitemOpen
\bibfield  {author} {\bibinfo {author} {\bibfnamefont {A.~W.}\ \bibnamefont
		{Glaetzle}}, \bibinfo {author} {\bibfnamefont {M.}~\bibnamefont {Dalmonte}},
	\bibinfo {author} {\bibfnamefont {R.}~\bibnamefont {Nath}}, \bibinfo {author}
	{\bibfnamefont {C.}~\bibnamefont {Gross}}, \bibinfo {author} {\bibfnamefont
		{I.}~\bibnamefont {Bloch}}, \ and\ \bibinfo {author} {\bibfnamefont
		{P.}~\bibnamefont {Zoller}},\ }\href {\doibase
	10.1103/PhysRevLett.114.173002} {\bibfield  {journal} {\bibinfo  {journal}
		{Phys. Rev. Lett.}\ }\textbf {\bibinfo {volume} {114}},\ \bibinfo {pages}
	{173002} (\bibinfo {year} {2015})}\BibitemShut {NoStop}%
\bibitem [{\citenamefont {Britton}\ \emph {et~al.}(2012)\citenamefont
	{Britton}, \citenamefont {Sawyer}, \citenamefont {Keith}, \citenamefont
	{Wang}, \citenamefont {Freericks}, \citenamefont {Uys}, \citenamefont
	{Biercuk},\ and\ \citenamefont {Bollinger}}]{Britton2012}%
\BibitemOpen
\bibfield  {author} {\bibinfo {author} {\bibfnamefont {J.~W.}\ \bibnamefont
		{Britton}}, \bibinfo {author} {\bibfnamefont {B.~C.}\ \bibnamefont {Sawyer}},
	\bibinfo {author} {\bibfnamefont {A.~C.}\ \bibnamefont {Keith}}, \bibinfo
	{author} {\bibfnamefont {C.-C.~J.}\ \bibnamefont {Wang}}, \bibinfo {author}
	{\bibfnamefont {J.~K.}\ \bibnamefont {Freericks}}, \bibinfo {author}
	{\bibfnamefont {H.}~\bibnamefont {Uys}}, \bibinfo {author} {\bibfnamefont
		{M.~J.}\ \bibnamefont {Biercuk}}, \ and\ \bibinfo {author} {\bibfnamefont
		{J.~J.}\ \bibnamefont {Bollinger}},\ }\href {\doibase 10.1038/nature10981}
{\bibfield  {journal} {\bibinfo  {journal} {Nature}\ }\textbf {\bibinfo
		{volume} {484}},\ \bibinfo {pages} {489} (\bibinfo {year}
	{2012})}\BibitemShut {NoStop}%
\bibitem [{\citenamefont {Wall}\ \emph {et~al.}(2015)\citenamefont {Wall},
	\citenamefont {Hazzard},\ and\ \citenamefont {Rey}}]{Wall2014}%
\BibitemOpen
\bibfield  {author} {\bibinfo {author} {\bibfnamefont {M.~L.}\ \bibnamefont
		{Wall}}, \bibinfo {author} {\bibfnamefont {K.~R.~A.}\ \bibnamefont
		{Hazzard}}, \ and\ \bibinfo {author} {\bibfnamefont {A.~M.}\ \bibnamefont
		{Rey}},\ }in\ \href {\doibase 10.1142/9789814678704_0001} {\emph {\bibinfo
		{booktitle} {From At. to Mesoscale}}},\ Vol.\ \bibinfo {volume} {80309}\
(\bibinfo  {publisher} {World Scientific},\ \bibinfo {year} {2015})\ pp.\
\bibinfo {pages} {3--37}\BibitemShut {NoStop}%
\bibitem [{\citenamefont {Schollw{\"{o}}ck}(2005)}]{Schollwock2005}%
\BibitemOpen
\bibfield  {author} {\bibinfo {author} {\bibfnamefont {U.}~\bibnamefont
		{Schollw{\"{o}}ck}},\ }\href {\doibase 10.1103/RevModPhys.77.259} {\bibfield
	{journal} {\bibinfo  {journal} {Rev. Mod. Phys.}\ }\textbf {\bibinfo {volume}
		{77}},\ \bibinfo {pages} {259} (\bibinfo {year} {2005})}\BibitemShut
{NoStop}%
\end{thebibliography}


\end{document}